\documentclass[aps,prd,preprintnumbers,superscriptaddress,nofootinbib,twocolumn]{revtex4-1}
\usepackage[dvipdfmx]{graphicx}
\usepackage{bm,latexsym,amsmath,amssymb,amsfonts,mathrsfs}
\usepackage{color}
\input{colordvi.tex}
\allowdisplaybreaks[1]
\usepackage[linktocpage=true]{hyperref}
\usepackage{url}
\hypersetup{
    colorlinks=true,
    citecolor=cyan,
    linkcolor=magenta,
}
\newcommand*{\D}{{\rm d}}

\newcommand{\fr}[2]{\frac{#1}{#2}}
\newcommand{\pa}{\partial}

\newcommand{\na}{\nabla}

\newcommand{\bra}[1]{\left( #1 \right)}
\newcommand{\brb}[1]{\left[ #1 \right]}
\newcommand{\brc}[1]{\left\{ #1 \right\}}
\newcommand{\be}{\begin{equation}}
\newcommand{\ee}{\end{equation}}
\newcommand{\bem}{\begin{bmatrix}}
\newcommand{\eem}{\end{bmatrix}}

\newcommand{\la}{\lambda}
\newcommand{\si}{\sigma}

\newcommand{\mn}{{\mu \nu}}
\newcommand{\mE}{\mathcal{E}}

\begin{document}
\title{Generalized 2D dilaton gravity and kinetic gravity braiding}
\author{Kazufumi~Takahashi}
\email[Email: ]{takahashik``at"rikkyo.ac.jp}
\affiliation{Department of Physics, Rikkyo University, Toshima, Tokyo 171-8501, Japan
}
\author{Tsutomu~Kobayashi}
\email[Email: ]{tsutomu``at"rikkyo.ac.jp}
\affiliation{Department of Physics, Rikkyo University, Toshima, Tokyo 171-8501, Japan
}
%
\begin{abstract}
We show explicitly that the nonminimal coupling between the scalar field and the
Ricci scalar in 2D dilaton gravity can be recast in the
form of kinetic gravity braiding (KGB). This is as it should be,
because KGB is the 2D version of the Horndeski theory.
We also determine all the static solutions
with a linearly time-dependent scalar configuration
in the shift-symmetric KGB theories in 2D.
\end{abstract}
\preprint{RUP-18-35}
\maketitle
\section{Introduction}

Dilaton gravity in 2D allows us to study aspects of
quantum gravity while avoiding technical complexity.
It arises from spherical or hyperbolic reduction of general relativity in 4D 
(or, more generically, Lovelock gravity in higher dimensions~\cite{Lovelock:1971yv}) 
as well as from string theory
(see Ref.~\cite{Grumiller:2002nm} for a review).
It would be very useful if one could
have the most general action involving
the metric and the scalar field with second-order
Euler-Lagrange (EL) equations and thus generalize the framework of 2D dilaton gravity.

In fact, such generalized 2D dilaton gravity has already been obtained earlier.
In Ref.~\cite{Horndeski:1974wa}, Horndeski derived the most general second-order
EL equations derived from a scalar-tensor theory
in four or less dimensions, together with the action whose variation yields those EL equations.
The action of the Horndeski theory in 2D spacetime
takes the form of {\it kinetic gravity braiding} (KGB)~\cite{Deffayet:2010qz}:
\begin{align}
S_{\rm KGB}=\int\D^2x\sqrt{-g}\left[K(\phi,X)-G(\phi,X)\Box\phi\right], \label{Horndeski2D}
\end{align}
where $K$ and $G$ are arbitrary functions of $\phi$
and $X:= -g^\mn \pa_\mu\phi\pa_\nu\phi/2$.

One may notice here that the familiar nonminimal coupling to
the Ricci scalar,
$\xi(\phi)R$,
is absent in the action~\eqref{Horndeski2D},
though it has been often considered in the literature.
For example,
Jackiw-Teitelboim gravity~\cite{Jackiw:1984je,Teitelboim:1983ux}
is described by the action of the form
\begin{align}
 S=\int\D^2x\sqrt{-g}\,\phi (R-\Lambda),
\end{align}
which can be embedded into Einstein-Maxwell-dilaton gravity in higher dimensions~\cite{Li:2018omr}.
More recently, the following action was studied
in Ref.~\cite{Kunstatter:2015vxa}: 
\begin{align}
S=\int\D^2x\sqrt{-g}&
\bigl[
\xi(\phi)R +k(\phi,X)
\notag \\ &
+C(\phi,X)\nabla^\mu\phi\nabla^\nu\phi \nabla_\mu \nabla_\nu\phi
\bigr],\label{MKaction}
\end{align}
which can describe the effective dynamics of spherically symmetric spacetime 
in Lovelock gravity~\cite{Kunstatter:2012jr,Kunstatter:2012kx} and was conjectured to be the most general 2D action
having second-order EL equations for the metric and the scalar field.
The second-order nature of the EL equations may imply the theory~\eqref{MKaction}
can be recast in the KGB form~\eqref{Horndeski2D} as the latter was shown to yield {\it the most general} second-order
EL equations, but this is far from obvious because we do not have a dictionary to translate
$(\xi,k,C)$ in~\eqref{MKaction} to $(K,G)$ in~\eqref{Horndeski2D}.
Thus, there is an apparent gap
between
existing 2D dilaton gravity and
the Horndeski theory in 2D,
concerning the nonminimal coupling to the Ricci scalar and
the terms involving a second derivative of $\phi$.

The main purpose of this short note is
to fill in this gap.
We demonstrate that $\xi(\phi)R$ [as well as the third term in~\eqref{MKaction}]
can be recast in the KGB form~\eqref{Horndeski2D}.
We also derive a new 2D black hole solution with
a linearly time-dependent scalar field in the presence of
shift symmetry.

\section{Nonminimal coupling and KGB}

In what follows, we use the notation~$\phi_\mu:= \na_\mu\phi$ and $\phi_\mn:= \na_\mu\na_\nu\phi$.
The EL equations in the 2D KGB theory
are obtained by taking variations of the action~\eqref{Horndeski2D}
with respect to $g_\mn$ and $\phi$:
	\begin{align}
	\mE^\mn:= \fr{1}{\sqrt{-g}}\fr{\delta S_{\rm KGB}}{\delta g_\mn}=0,~~~\mE_\phi:= \fr{1}{\sqrt{-g}}\fr{\delta S_{\rm KGB}}{\delta \phi}=0.
	\end{align}
The explicit form for $\mE^\mn$ is
	\begin{align}
	\mE^\mn&=\fr{1}{2}K_X\phi^\mu\phi^\nu+\fr{1}{2}Kg^\mn-\phi^{(\mu}\na^{\nu)}G \nonumber \\
	&~~~+\fr{1}{2}g^\mn \phi^\la\na_\la G-\fr{1}{2}G_X\phi^\mu\phi^\nu\Box\phi \nonumber \\
	&=XG_X\bra{g^\mn\Box\phi-\phi^\mn}+\fr{1}{2}\bra{K_X-2G_\phi}\phi^\mu\phi^\nu \nonumber \\
	&~~~+\fr{1}{2}(K-2XG_\phi)g^\mn, \label{metricEL}
	\end{align}
where subscripts~$\phi$ and $X$ represent partial derivatives with respect to $\phi$ and $X$, respectively.
Here, we have used the identity
	\begin{align}
	&g^\mn\phi^\la\phi^\si\phi_{\la\si}+\phi^\mu\phi^\nu\Box\phi
	-2\phi^\la\phi_\la^{(\mu}\phi^{\nu)} \nonumber \\
	&=2X\bra{\phi^\mn-g^\mn\Box\phi}, \label{id2D}
	\end{align}
which holds in two spacetime dimensions.
This identity can be checked on a component-by-component basis by choosing a coordinate system where the metric is conformally flat, which is always possible in 2D.
Regarding the scalar-field equation of motion, ${\cal E}_\phi$ satisfies
	\begin{align}
	\mE_\phi\phi^\nu=2\,\na_\mu \mE^\mn, \label{NI}
	\end{align}
which is nothing but the Noether identity associated with general covariance.
Therefore, $\mE_\phi=0$ is automatically satisfied for a configuration~$(g_\mn,\phi)$ that satisfies $\mE^\mn=0$.

Now we consider a nonminimal coupling between the scalar field and the Ricci scalar in 2D spacetime:
	\begin{align}
	S_{\rm NMC}=\int \D^2x\sqrt{-g}\xi(\phi)R. \label{nonmini}
	\end{align}
It is obvious that the action yields second-order EL equations and thus should be recast in the form of the KGB action~\eqref{Horndeski2D}.
We verify this by comparing the EL equation for the metric derived from the action~\eqref{nonmini} with Eq.~\eqref{metricEL}.
Varying the action~\eqref{nonmini} with respect to $g_\mn$, we have
	\begin{align}
	\fr{1}{\sqrt{-g}}\fr{\delta S_{\rm NMC}}{\delta g_\mn}&=\na^\mu\na^\nu
  \xi(\phi)-g^\mn\Box \xi(\phi) \nonumber \\
	&=\xi'(\phi)\phi^\mn-\xi'(\phi)g^\mn\Box\phi \nonumber \\
	&~~~+\xi''(\phi)\phi^\mu\phi^\nu+2\xi''(\phi)Xg^\mn,
	\end{align}
where a prime stands for differentiation with respect to $\phi$.
The right-hand side is reproduced from Eq.~\eqref{metricEL} by taking
	\begin{align}
	K=2\xi''(\phi)X(2-\ln X),\quad G=-\xi'(\phi)\ln X. \label{nonmini2H}
	\end{align}
Note that this choice also reproduces the corresponding scalar-field EL equation thanks to the identity~\eqref{NI}.
Thus, we have established that the nonminimal coupling to the Ricci scalar described by the action~\eqref{nonmini} is written in the KGB form,
and hence is redundant in 2D.

The discussion so far may remind us of the nonminimal coupling
to the Gauss-Bonnet term in 4D,
$\xi(\phi)\times $(Gauss-Bonnet term).
Though we know that the Horndeski theory gives the most general second-order
scalar-tensor theory in 4D, it is far from trivial that this term
is indeed included in the Horndeski action.
However, by comparing the EL equations directly, Ref.~\cite{Kobayashi:2011nu}
showed how one can express the nonminimal coupling to the Gauss-Bonnet term
in terms of the four functions of $(\phi,X)$ in the Horndeski action in 4D.
Notice that the Gauss-Bonnet term is topological in 4D, and so is
the Ricci scalar in 2D.
Thus, the situation in 2D dilaton gravity is
analogous to that in 4D gravity involving the Gauss-Bonnet term.

Note in passing that the third term in the action~\eqref{MKaction}
can also be rewritten in the KGB form in the following way.
For $C(\phi,X)$ in~\eqref{MKaction} let us define
\begin{align}
D(\phi,X):=\int^XC(\phi,X')\D X'.
\end{align}
Here, $D$ is determined only up to an arbitrary function of $\phi$, but this ambiguity is irrelevant in the following argument.
Then, we have
\begin{align}
\phi^\mu\nabla_\mu D = -2 D_\phi X - C\phi^\mu\phi^\nu\phi_{\mu\nu}.
\end{align}
Integration by parts leads to
\begin{align}
\int\D^2x \sqrt{-g}C\phi^\mu\phi^\nu\phi_{\mu\nu}
=\int\D^2x\sqrt{-g}\bra{-2D_\phi X+D\Box\phi},
\label{transformKGB}
\end{align}
and now the right-hand side is of the KGB form.
Note that the formula~\eqref{transformKGB} holds in any dimensions.

We have thus seen
that the action~\eqref{MKaction} studied in Ref.~\cite{Kunstatter:2015vxa}
reduces to the most general second-order scalar-tensor theory
in 2D derived by Horndeski~\cite{Horndeski:1974wa}.
Summarizing the above results, the former characterized by $k(\phi,X)$, $C(\phi,X)$, and $\xi(\phi)$ is equivalent to the latter with
	\be
	\begin{split}
	K&=k-2D_\phi X+2\xi''(\phi)X(2-\ln X), \\
	G&=-D-\xi'(\phi)\ln X,
	\end{split} \label{MK2KGB}
	\ee
where $D$ is defined so that $D_X=C$.

More generically, the quartic Galileon Lagrangian~\cite{Deffayet:2011gz} in 2D,
	\begin{align}
	S_{\rm Gal}=\int \D^2x\sqrt{-g}\brc{F(\phi,X)R+F_X\brb{(\Box\phi)^2-\phi_\mn\phi^\mn}}, \label{nonminiX}
	\end{align}
can be recast in the KGB form~\eqref{Horndeski2D}.
To see this, the following identity is useful:
	\begin{align}
	&FR+F_X\brb{(\Box\phi)^2-\phi_\mn\phi^\mn}+\na_\mu\brb{\fr{F}{X}\bra{\phi_\nu\phi^\mn-\phi^\mu\Box\phi}} \nonumber \\
	&=F_\phi\bra{2\Box \phi+\fr{1}{X}\phi^\mu\phi^\nu\phi_\mn},
	\end{align}
which can be verified by using Eq.~\eqref{id2D} and $R_\mn=\fr{1}{2}Rg_\mn$.
Thus, the action~\eqref{nonminiX} is written as
	\begin{align}
	S_{\rm Gal}=\int \D^2x\sqrt{-g}F_\phi\bra{2\Box \phi+\fr{1}{X}\phi^\mu\phi^\nu\phi_\mn}, \label{nonminiX2}
	\end{align}
and this reduces to the KGB form using the formula~\eqref{transformKGB}.

\section{All static spacetimes with a linearly time-dependent scalar field in shift-symmetric theories}

Now let us derive static spacetimes
in the 2D KGB theory in the presence of
the shift symmetry, $\phi\to \phi+c$, where $c$ is a constant.
In this case, $K$ and $G$ are 
functions of $X$ only,
which greatly simplifies the analysis.
Note that the nonminimal coupling of the form $\phi R$
respects the shift symmetry, because $R$ is a total divergence in 2D.
Indeed, $\phi R$ corresponds to
$K=0$ and $G=-\ln X$
in the KGB language. Therefore, the simplest theory in the shift-symmetric class with the term~$\phi R$
would be given by
\begin{align}
S&=\int \D^2x\sqrt{-g}\left(\phi R+X-2\Lambda\right)
\notag \\ &=\int \D^2x\sqrt{-g}\left[X-2\Lambda+(\ln X)\Box\phi\right], \label{simplest}
\end{align}
where $\Lambda$ is a constant.
A black hole solution in this particular theory was
obtained in Refs.~\cite{Mann:1991md,Mureika:2011py}.
In this section, we extend the previous results to
derive the static metric in the most general shift-symmetric KGB theory
while allowing for the linear time-dependence of the scalar field.

The most general form of the metric of 2D static spacetime is
given by
\begin{align}
\D s^2=-f(x)\D t^2+\frac{\D x^2}{f(x)}.\label{ansatz_metric}
\end{align}
In shift-symmetric theories, the scalar field
in static spacetime may
depend linearly on time because the EL equations
contain $\phi_\mu$ and $\phi_{\mu\nu}$ but not $\phi$
without derivatives.
We therefore look for
solutions of the form
\begin{align}
\phi=qt+\psi(x),\label{ansatz_scalar}
\end{align}
where $q$ is a nonvanishing constant.
We have
\begin{align}
X=\frac{1}{2}\left[\frac{q^2}{f}-(\psi')^2f\right],
\end{align}
where a prime here denotes differentiation with respect to $x$.

The EL equations read ${\cal E}_{\mu\nu}=0$, where
\begin{align}
{\cal E}_{tt}&=\frac{1}{2} \left[
K_Xq^2-Kf-G_X X f\left(2f\psi''+f'\psi'\right)\right],\label{ett}
\\
{\cal E}_{tx}&=\frac{q}{2}\left(K_X\psi'+G_{X}X\frac{f'}{f}\right),\label{etr}
\\
{\cal E}_{xx}&=\frac{1}{2}\left[K_X(\psi')^2+\frac{K}{f}
+G_XX\frac{f'}{f}\psi'\right].\label{err}
\end{align}
Since the equation of motion for $\phi$
is automatically satisfied
if ${\cal E}_{\mu\nu}=0$ is satisfied,
we do not need the explicit expression for ${\cal E}_\phi$.
Interestingly,
one can integrate Eqs.~\eqref{ett}--\eqref{err}
to obtain the most general solution of the form~\eqref{ansatz_metric}
and~\eqref{ansatz_scalar}
without assuming the concrete form of the functions $K(X)$ and $G(X)$.

From ${\cal E}_{tx}=0$ and ${\cal E}_{xx}=0$,
we obtain
\begin{align}
\frac{K(X)}{f}=0.\label{KX0}
\end{align}
This means that $X$ must be a constant,
\begin{align}
X=X_0,
\end{align}
where $X_0$ is a real root of $K(X)=0$ (if it exists)
and hence is determined from the Lagrangian under consideration.
Then, using ${\cal E}_{tx}=0$ we get
\begin{align}
\psi'=\lambda\frac{f'}{f},\label{psidash=}
\end{align}
where
\begin{align}
\lambda:=-\frac{X_0G_X(X_0)}{K_X(X_0)}
\end{align}
is a constant.
Here and hereafter we assume that
$K_X(X_0)\neq 0$ and $\lambda \neq 0$.
Equation~\eqref{psidash=} immediately gives
\begin{align}
\phi =q t+ \lambda \ln f +{\rm const},\label{sol_scalar}
\end{align}
where the integration constant may be absorbed into
a trivial shift of the origin of $t$.

Using Eqs.~\eqref{KX0} and~\eqref{psidash=}, the EL equation
${\cal E}_{tt}=0$ reduces to
\begin{align}
q^2+\lambda^2\left[2ff''-(f')^2\right]=0.
\end{align}
This equation can be solved to give
\begin{align}
f=C-\mu x-\frac{q^2-\lambda^2\mu^2}{4\lambda^2C}x^2,\label{sol_f}
\end{align}
where $C(\ne 0)$ and $\mu$ are integration constants.
One can verify that $X$ is indeed a constant for the above configurations of $\phi$ and $f$:
\begin{align}
X=\frac{q^2-\lambda^2\mu^2}{2C}\;\left(=X_0 \right),
\end{align}
which leads to
\begin{align}
q=\pm\left(\lambda^2\mu^2+2CX_0\right)^{1/2}.
\end{align}
In order for this to be a real number,
the two integration constants must satisfy
$\lambda^2\mu^2\ge -2CX_0$.

We have thus obtained the general static solution
with a linearly time-dependent scalar field,~\eqref{sol_scalar}
and~\eqref{sol_f}, which is characterized by
the two integration constants, $C$ and $\mu$.
Interestingly, the structure of~\eqref{sol_f}
is the same as the metric function of a 2D black hole with a cosmological constant
found in the literature (see Ref.~\cite{Mureika:2011py}).
As we will see shortly, $\mu$ is related to the mass of the black hole.
The effective cosmological constant is given by
$\Lambda_{\rm eff}=(q^2-\lambda^2\mu^2)/(4\lambda^2C)=X_0/(2\lambda^2)$.
Note that $\Lambda_{\rm eff}$ is determined from the functions in
the Lagrangian and it is independent of the integration constants.
For the simplest model~\eqref{simplest}, one has $\Lambda_{\rm eff}=\Lambda$. 
Note also that the above solution has a curvature singularity at $x=0$ unless $\mu=0$.
It is possible to obtain a nonsingular black hole by introducing appropriate terms that break the shift symmetry (see, e.g., Refs.~\cite{Trodden:1993dm,Kunstatter:2015vxa}).

Let us now consider a point mass $M$ placed at $x=0$.
The EL equation is then given by
${\cal E}_{\mu\nu}=-T_{\mu\nu}/2=-(M/2)\delta(x)\delta_\mu^0\delta_\nu^0$.
The gravitational field sourced by this point mass is
described by
\begin{align}
f=C-\mu |x|-\Lambda_{\rm eff}x^2.\label{sol_f0}
\end{align}
Indeed, one finds that
\begin{align}
{\cal E}_{tt}=-2K_X(X_0)\lambda^2 C \mu\,\delta(x),
\end{align}
showing that $\mu$ is proportional to the mass.
See Ref.~\cite{Christensen:1991dk} for the causal structure
of spacetime described by~\eqref{sol_f0}.

So far we have assumed that $q\neq 0$.
If we start with the ansatz $\phi = \psi(x)$,
the $(tx)$ equation becomes trivial
and as a result the EL equations would admit more variety of solutions,
depending on the concrete form of the functions $K(X)$ and $G(X)$.
Note that even in that case our solution remains a solution,
and therefore
taking $q=0$ in our solution does make sense:
\begin{align}
f=C\left(1-\frac{\mu |x|}{2C}\right)^2,
\quad
\phi = \lambda\ln f,
\quad
C=-\frac{\lambda^2\mu^2}{2X_0}.
\end{align}
We see that for $q=0$ the two integration constants are related to each other,
and
the solution has a degenerate horizon (when $X_0<0$ and $\mu>0$).

\section{Conclusions}

In this short note, we have clarified
how the action~\eqref{MKaction} with
the nonminimal coupling to the Ricci scalar, $\xi(\phi)R$,
can be expressed in the form of kinetic gravity braiding (KGB) in 2D,
i.e.,
the most general 2D scalar-tensor theory having second-order
EL equations.
The relation between these two theories is explicitly given by~\eqref{MK2KGB}.

We have also obtained all static spacetimes with
a linearly time-dependent scalar field in the shift-symmetric 2D KGB theory
without assuming any specific form of the functions in the action.
For a black hole solution,
it would be intriguing to study its thermodynamical properties,
but this point is beyond the scope of the present note.
Moreover, it would also be interesting to construct nonsingular black holes in the 2D KGB theory, which could give us some hints on nonsingular black holes in 4D.

It is interesting to mention another characteristic of
the 2D scalar-tensor theory.
In 4D, by means of metric redefinition of the form~$g_\mn\to A(\phi,X)g_\mn+B(\phi,X)\phi_\mu \phi_\nu$
(called {\it disformal transformation}~\cite{Bekenstein:1992pj}), one can promote the Horndeski class
to a broader one~\cite{Crisostomi:2016czh,Achour:2016rkg,BenAchour:2016fzp}.
However, this is not the case in 2D:
The 2D Horndeski theory (i.e., the KGB theory) is closed under the disformal transformation.
This fact makes the KGB theory a firm ground for studying 2D dilaton gravity.

Yet another is that the Horndeski theory can be applied to the AdS/CFT correspondence in higher dimensions~\cite{Li:2018kqp,Li:2018rgn}.
Thus, it may also be interesting to explore an application of the KGB theory to the AdS${}_2$/CFT${}_1$ correspondence.

Finally, we would like to comment on 2D dilaton gravity as an effective theory to describe spherical (or hyperbolic) spacetime in higher dimensions.
As mentioned earlier, a spherical reduction of Lovelock gravity leads to the action of the form~\eqref{MKaction}, whose EL equations are at most of second order.
This is as expected because Lovelock gravity is the most general metric theory having second-order EL equations in arbitrary dimensions.
However, there are some examples of theories having higher derivatives in their EL equations but still having only up to second derivatives under the spherically symmetric ansatz~\cite{Oliva:2010eb,Oliva:2010zd,Myers:2010ru}.
From these observations, it would be intriguing to identify all such higher derivative theories whose spherical reduction yields second-order field equations. 
This issue will be addressed in a future publication.

\acknowledgments
We are grateful to
Shunichiro Kinoshita, Hideki Maeda,
and Keisuke Nakashi for enlightening conversations.
The work of KT was supported by
JSPS Research Fellowships for Young Scientists No.~17J06778.
The work of TK was supported by
MEXT KAKENHI Grant Nos.~JP15H05888, JP17H06359, JP16K17707, JP18H04355,
and MEXT-Supported Program for the Strategic Research Foundation at Private Universities,
2014-2018 (S1411024).

\bibliography{2d}
\bibliographystyle{JHEP.bst}
\end{document}